\documentclass{ws-procs975x65}
\usepackage{graphicx}

\def\beq{\begin{equation}}
\def\eeq{\end{equation}}

\def\figsubcap#1{\par\noindent\centering\footnotesize(#1)}

\begin{document}

\title{Modified Dark Matter} 
\author{Y.\ Jack Ng$^*$}
\address{Institute of Field Physics, Department of Physics and
Astronomy,\\
University of North Carolina, Chapel Hill, NC 27599, U.S.A.\\
$^*$E-mail: yjng@physics.unc.edu}
 
\author{Doug Edmonds}
\address{Department of Physics, Emory \& Henry College, Emory, VA 24327, 
U.S.A.\\
E-mail: dedmonds@ehc.edu}

\author{Duncan Farrah, Djordje Minic and Tatsu Takeuchi}
\address{Department of Physics, Virginia Tech, Blacksburg, VA 24061, U.S.A.\\
E-mail:farrah@vt.edu, dminic@vt.edu, takeuchi@vt.edu}

\author{Chiu Man Ho}
\address{Department of Physics and Astronomy, Michigan State University,\\ 
East Lansing, MI 48824, U.S.A.\\
E-mail: cmho@msu.edu}



\begin{abstract}
Modified dark matter (MDM, formerly known as MoNDian dark matter) is a 
phenomenological model of dark matter, inspired by quantum gravity.   
We review the construction of MDM by generalizing entropic
gravity to de-Sitter space as is appropriate for an accelerating universe (in 
accordance with the $\Lambda$CDM model). Unlike cold dark matter models,
the MDM mass profile depends on the baryonic mass.
We successfully fit the rotation curves to a sample of 30 local spiral 
galaxies with a single free parameter 
(viz., the mass-to-light ratio for each galaxy). 
We show that dynamical
and observed masses agree in a sample of 93 galactic clusters.  
We also comment on strong gravitational lensing in the context of MDM.
\end{abstract}


\bodymatter

\section{Introduction and Summary}

The cold dark matter (CDM) model 
successfully explains several astrophysical phenomena. 
These include flat galactic rotation curves, gravitational lensing,
elemental abundances from big bang nucleosynthesis,
and the power spectrum of cosmic microwave background anisotropies. 
This consistency has led to the widespread acceptance of the $\Lambda$CDM 
paradigm,
in which the Universe also exhibits a cosmological constant $\Lambda$.
CDM does, however, have remaining tensions with observations, especially on 
$\lesssim$ Mpc scales.
These include inconsistency with the observed asymptotic velocity-mass ($v^4 
\propto M$) scaling in the Tully-Fisher relation \cite{TF}. 

Efforts have been made to construct theories that better match observations on 
galactic length scales than CDM. The most prominent of these is modified 
Newtonian dynamics (MOND) \cite{mond} 
proposed by Milgrom. In MOND the equation of motion is 
$F=ma\mu(x)$, such that the smooth interpolating function
$\mu(x) = 1, x$ for $x\gg 1$ and $x\ll 1$ respectively,
where $x \equiv a/a_c$ with the critical acceleration $a_c$ 
being a tunable parameter,
found to be numerically related to the speed of light $c$ and the Hubble 
parameter $H$ as
$a_c \approx c \,H/(2\, \pi) \sim 10^{-8} \mathrm{cm/s^2}$. 
A favorite interpolating function is given by
$\mu(x) = \dfrac{x}{(1+x^2)^{1/2}}$ which we will use in section 4.
For a given source mass $M$, we have $a=a_N \equiv GM/r^2$ 
when $a\gg a_c$, while $a=\sqrt{a_c a_N}$ when $a\ll a_c$,
where $a_N$ is the usual Newtonian acceleration without dark matter.
It is readily shown that on the outer edges of a galaxy where 
gravity is weak, MOND yields asymptotically flat rotation curves, and 
$v^4\propto M$, the Tully-Fisher relation as observed.
MOND however struggles to reproduce observations at cluster and cosmological 
length scales \cite{bullet}.
Hence, CDM is usually preferred over MOND, with efforts ongoing to reconcile 
CDM with observations on $\lesssim$ Mpc scales.

Modified dark matter (MDM) \cite{Ho1} is a new form of dark matter quantum that 
theoretically 
behaves like cold dark matter (CDM) at 
cluster and cosmic scales, and naturally 
accounts for the scaling usually associated with MOND at the galactic scales.
The latter feature explains why MDM was originally known as MONDian dark matter
\cite{Ho1};
but in view of the confusing connotation to which this nomenclature has 
probably 
given rise and in order not to distract from the fact that the model {\it is}
a model of dark matter, we have settled on the name ``modified dark matter" 
(to distinguish it from cold dark matter.)  MDM is a
phenomenlogical model inspired by quantum gravity.  We recall its construction
in section 2.  In section 3 we provide the first observational test of MDM by 
fitting rotation curves to a sample of 30 local spiral galaxies ($z \approx 
0.003$) \cite{Edm1} and we show that MDM is a more economical model than CDM. 
We test the MDM model at the cluster scale \cite{Edm1} 
in section 4 where we show that
MDM is superior to MOND.  In both the galactic and cluster scales, MDM fares
well.  Future work will include:
1. Gravitational lensing (can it distinguish MDM from CDM?)
2. Interactions of MDM 
(unusual particle phenomenology? 
The Bullet 
Cluster; how strongly
coupled is MDM to baryonic matter and how does MDM self-interact?)
3. Tests at cosmic scales 
(acoustic oscillations measured in the CMB,
simulations of structure formation?)

\section{Constructing MDM}

MDM is a phenomenological model of dark matter inspired by quantum gravity,
based on a simple generalization of E. Verlinde's recent proposal 
of entropic gravity \cite{verlinde},
which happens to provide
a convenient framework for its construction \cite{Ho1}.
Consider a particle with mass $m$ approaching a holographic screen
at temperature $T$.  Using the first law of thermodynamics to 
introduce the concept of entropic force
$
F = T \frac{\Delta S}{\Delta x},
$
and invoking Bekenstein's original arguments \cite{bekenstein}
concerning the entropy $S$ of black holes,
$\Delta S = 2\pi k_B \frac{mc}{\hbar} \Delta x$,
we get $ F = 2\pi k_B \frac{mc}{\hbar} T$. 
In a deSitter space with cosmological constant $\Lambda$, the net 
Unruh-Hawking temperature, \cite{unruh}
as measured by 
a non-inertial observer with acceleration $a$ 
relative to an inertial observer, is 
$T = \frac{\hbar}{2\pi k_B c} [\sqrt{a^2+a_0^2} - a_0]$, \cite{deser}
where $a_0 \equiv \sqrt{\Lambda / 3}$.  Hence the
entropic force (in deSitter space) is given by 
$
F =  m [\sqrt{a^2+a_0^2}-a_0].
$
For $ a \gg a_0$, we have $F/m \approx a$ which gives $a = 
a_N 
$, the familiar Newtonian value for the acceleration 
due to the source $M$. But for $a \ll a_0$, 
$F \approx m \frac{a^2}{2\,a_0},$ so
the terminal velocity $v$ of the test mass $m$ in a circular motion 
with radius $r$ should be determined from
$ m a^2/(2a_0) = m v^2/r$.  In this small acceleration regime,
in order to fit the galactic rotation curves as Milgrom did, we 
require $F \approx m \sqrt{a_N a_c}\,$, which, in turn, requires
$ a \approx \left( 4a_N \,a_0^2 a_c \, \right)^{\frac14}
\approx \left( 2 a_N \,a_0^3 / \pi \, \right)^{\frac14}$, 
where we have noted that numerically $a_0 \approx 2 \pi a_c$.
From our perspective, MoND is a {\it classical} phenomenological
consequence of {\it quantum} gravity (with the $\hbar$ dependence
in $T \propto \hbar$ and $S \propto 1/\hbar$ cancelled out). 
\cite{Ho1}

Having generalized Newton's 2nd law, we \cite{Ho1}
can now follow the second half of Verlinde's argument \cite{verlinde} 
to generalize Newton's law of gravity
$a= G M /r^2$, by considering an
imaginary quasi-local (spherical) holographic screen of area $A=4 \pi
r^2$ at temperature $T$.  Invoking the 
equipartition of energy $E= \frac{1}{2} N k_B T$
with $N = Ac^3/(G \hbar)$ being 
the total number of degrees of freedom (bits) on the screen, 
as well as the Unruh
temperature formula and the fact that $E= M_{total} c^2$, we
get 
$2 \pi k_B T 
= G\,M_{total} /r^2$,
where $M_{total} = M + M'$ represents the \emph{total} mass 
enclosed 
within the volume $V = 4 \pi r^3 / 3$, with
$M'$ being some unknown mass, i.e., dark
matter.  For $a \gg a_0$, consistency with the Newtonian force law 
$a \approx a_N$ implies $M' \approx 0$.  But
for $a \ll a_0$, consistency with the condition 
$a \approx \left( 2 a_N \,a_0^3 / \pi \right)^{\frac14}$ requires 
$M' \approx \frac{1}{\pi} \left(\,\frac{a_0}{a}\,\right)^2\, M$
whence it follows that
$
F = m a_N \left[ 1 + \frac{1}{\pi} \left( \frac{a_0}{a} \right)^2 \right].
$
Thus dark matter indeed exists!
And the MOND {\it force law} derived above, at the galactic scale, 
{\it is simply a manifestation of dark matter}! 


\section{Fitting galactic rotation curves with MDM mass profiles}

In order to test MDM with galactic rotation curves, we fit computed rotation 
curves to a selected sample of Ursa Major galaxies given in Ref. 10.
The sample contains both high surface brightness (HSB) and 
low surface brightness (LSB) galaxies. 
The rotation curves, predicted by MDM as given above by 
$
F =  m [\sqrt{a^2+a_0^2}-a_0]
  = m a_N \left[ 1 + \frac{1}{\pi} \left( \frac{a_0}{a} \right)^2 \right]
$
along with $ F = m v^2 / r$ for circular orbits, can be
solved for $a(r)$ and $v(r)$. We \cite{Edm1}
fit these to the observed rotation curves as determined in Ref. 10,
using a least-squares fitting routine. As in Ref. 10,
the mass-to-light ratio $M/L$, 
which is our {\it only} fitting parameter for MDM, is assumed constant for a 
given galaxy but allowed to vary between galaxies. Once we have $a(r)$, we 
can find the MDM density profile by using 
$M' \approx \frac{1}{\pi} \left(\,\frac{a_0}{a}\,\right)^2\, M$
to give
$
\rho'(r) \;=\; \left( \frac{a_c}{r} \right)^2 \frac{d}{dr} \left( \frac{M}{a^2} 
\right).
$

\begin{figure}[h]%
\begin{center}
\parbox{2.3in}{\includegraphics[angle=90,width=1.8in]{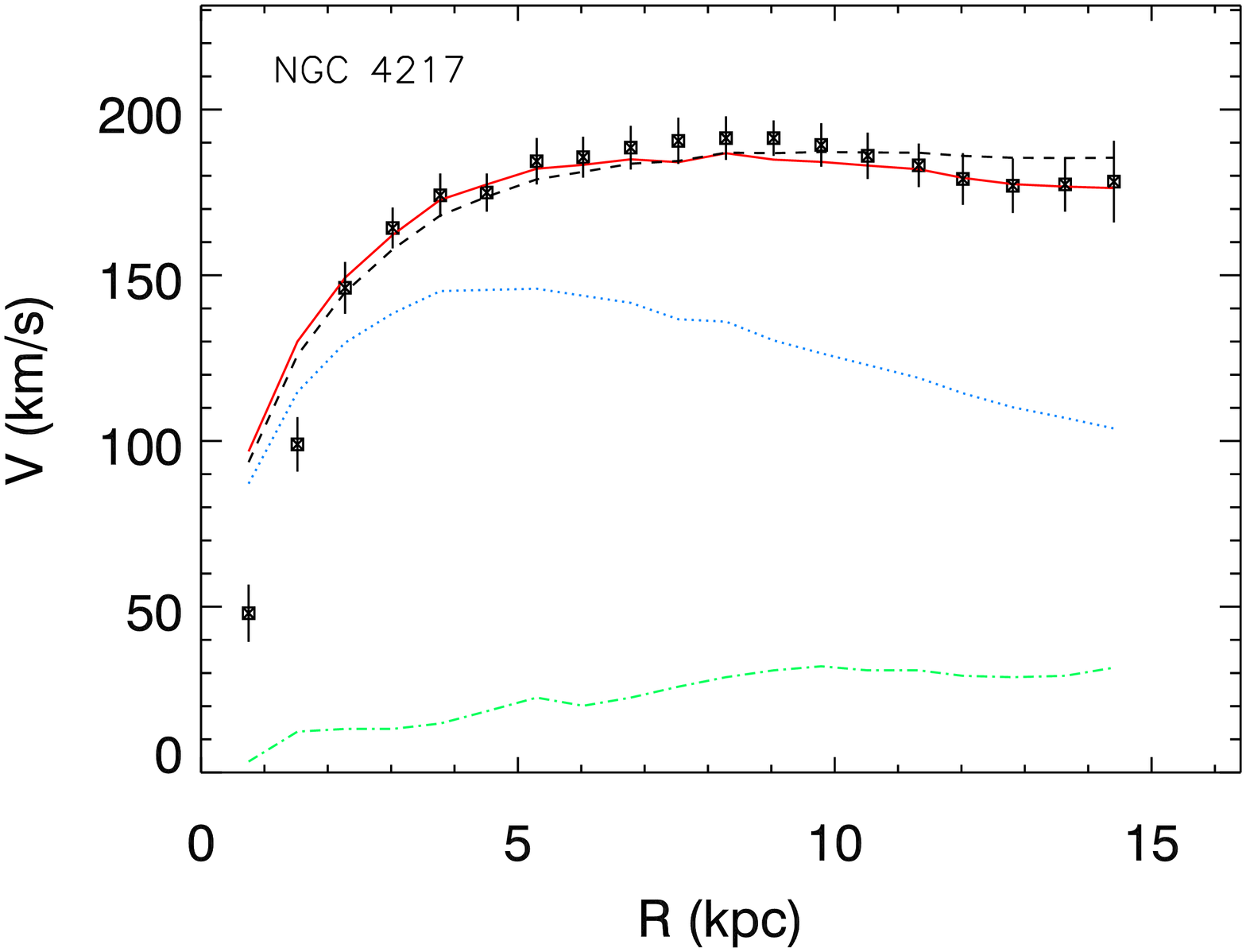} 
\figsubcap{a}}
\hspace*{4pt}
\parbox{2.3in}{\includegraphics[angle=90,width=1.8in]{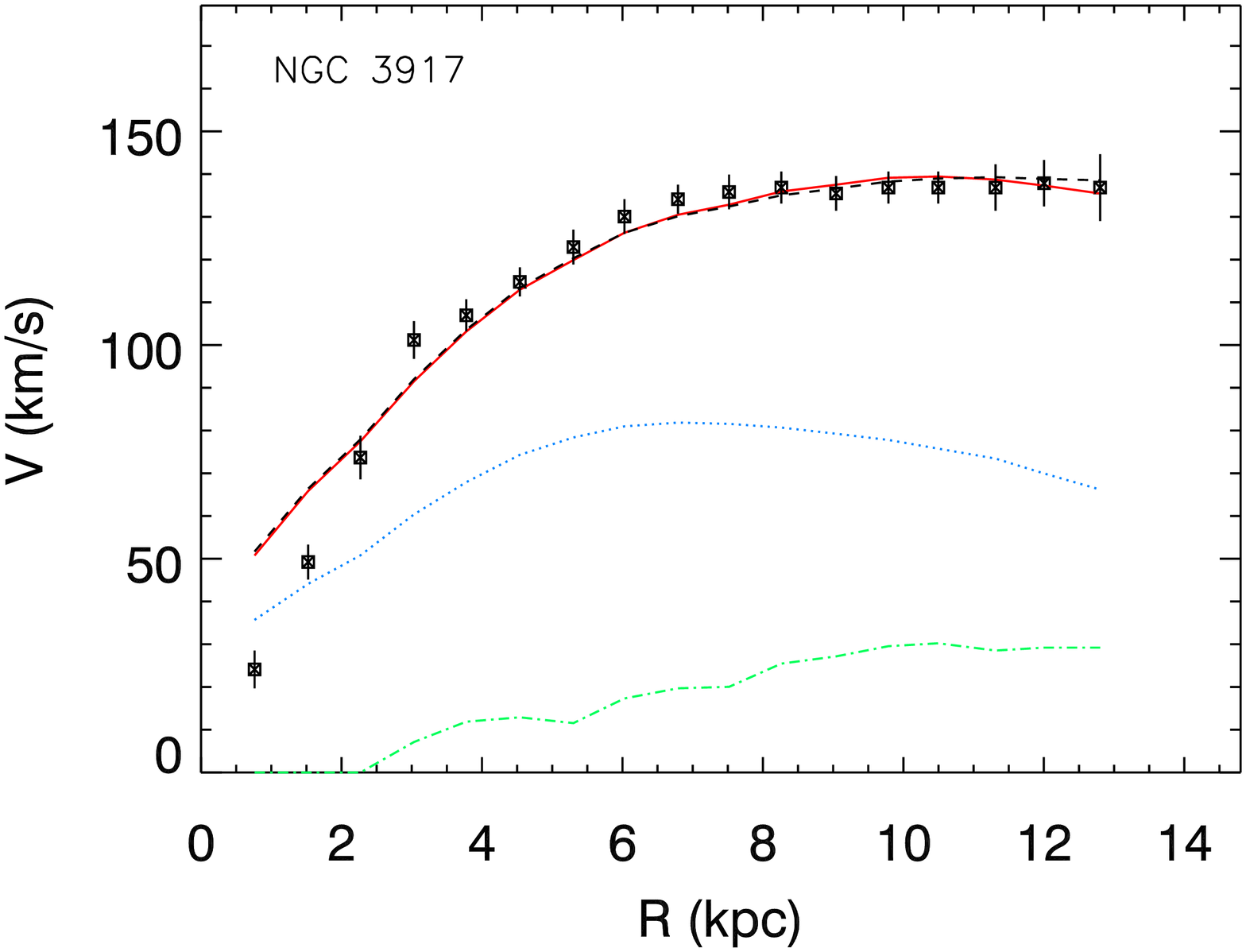} 
\figsubcap{b}}
\caption{Galactic rotation curves:
(a) NGC 4217 (HSB); (b) NGC 3917 (LSB).}
\label{fig1.2}
\end{center}
\end{figure}

Rotation curves predicted by MDM for NGC 4217, a typical HSB galaxy, 
and NGC 3917, a typical LSB galaxy in the sample
are shown in Fig. ~\ref{fig1.2}. 
(See Ref. 5
for the rotation curves for the other 28 galaxies.)

In these figures, observed 
rotation curves are depicted as filled circles with error bars, and for the 
two curves at the bottom, the dotted 
and dash-dotted lines show the stellar and interstellar gas rotation curves, 
respectively. The solid lines and dashed lines are rotation curves predicted by 
MDM and the standard cold dark matter (CDM) paradigm respectively. 
For the CDM fits, we use the Navarro, Frenk \& White (NFW) \cite{nfw} 
density profile, employing {\it three} free parameters (one of which is
the mass-to-light ratio.)  It is fair
to say that both models fit the data well; \footnote{We should point out that
the rotation curves predicted by MDM and MOND 
have been found \cite{Edm1} to be virtually indistinguishable
over the range of observed radii and both employ only 1 free parameter.}
but we remind the readers that while the MDM fits use only 1 free parameter,
for the CDM fits one needs to use 3 free parameters.  Thus the MDM model is a 
more
economical model than CDM in fitting data at the galactic scale.\footnote
{Since MDM employs only the minimal number of free parameters, the
speaker (YJN) called it the ``minimal dark matter" model in his talk at MG14. 
But 
later he found out that the name had been used before for another model.  His
collaborators and he decided to change the name to ``modified dark matter".}

\begin{figure}[h]%
\begin{center}
\parbox{2.2in}{\includegraphics[angle=90,width=1.8in]{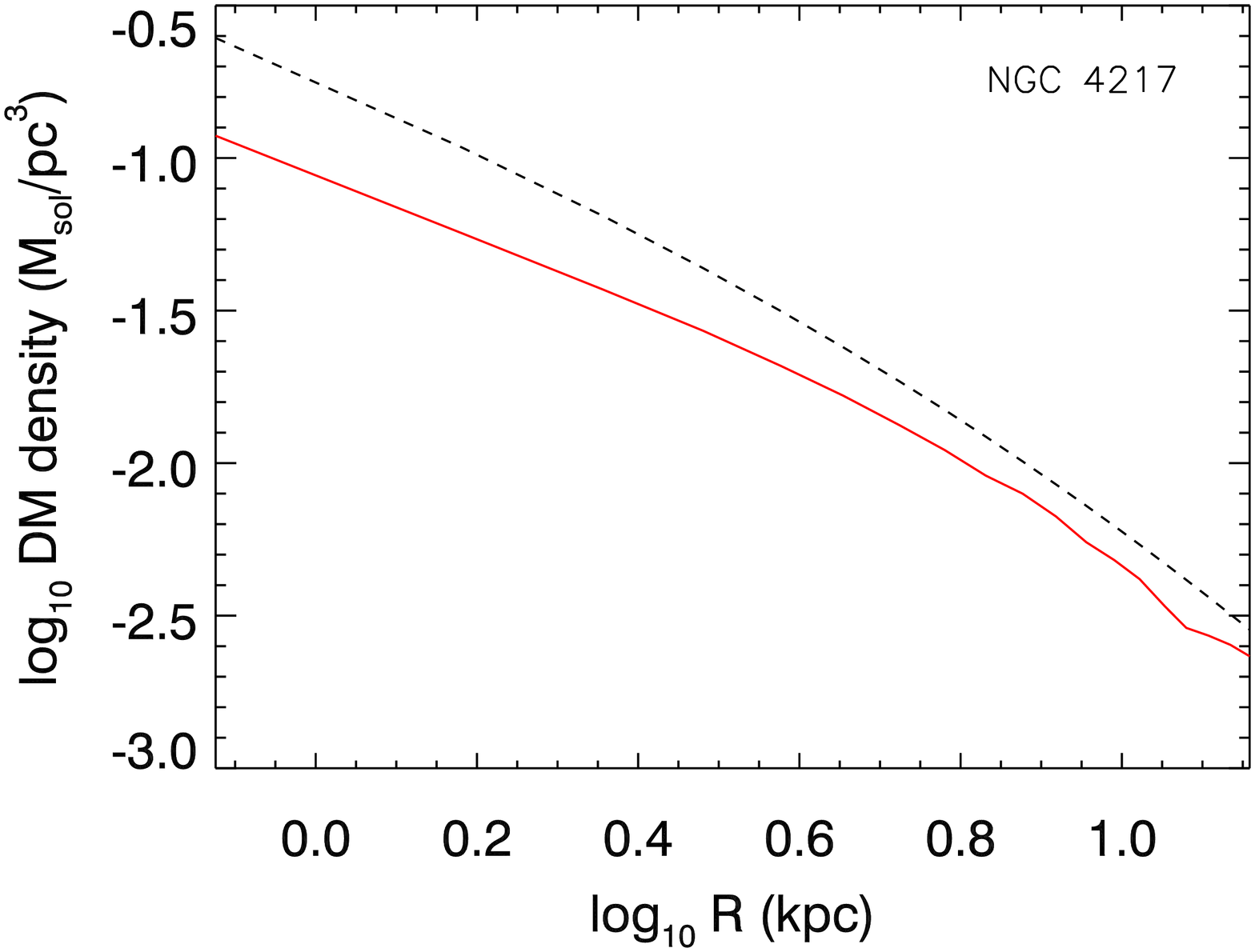}
\figsubcap{a}}
\hspace*{4pt}
\parbox{2.2in}{\includegraphics[angle=90,width=1.8in]{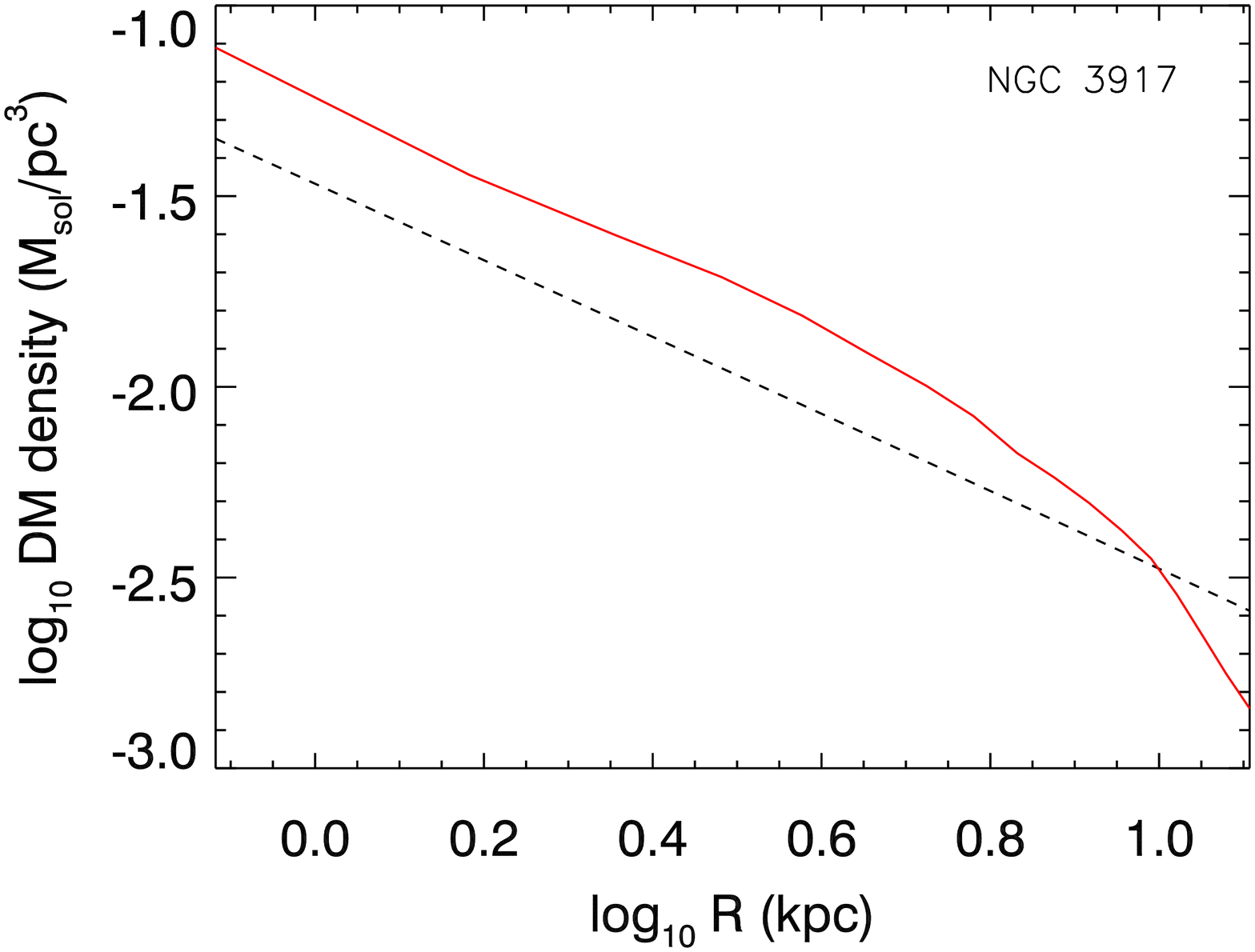}
\figsubcap{b}}
\caption{Dark matter density profiles:
(a) NGC 4217 (HSB); (b) NGC 3917 (LSB).}
\label{fig3.4}
\end{center}
\end{figure}


Shown in Fig. ~\ref{fig3.4} 
are the dark matter density
profiles predicted by MDM (solid lines) and CDM (dashed lines) for
the HSB galaxy NGC 4217 and the LSB galaxy NGC 3917 in the 
sample respectively.

\section{Testing MDM with Galactic Clusters}

To test MDM with astronomical 
observations at a larger scale, we \cite{Edm1}
compare dynamical and observed masses in a large sample of 
galactic clusters. First, let us recall that
the MDM profile 
$M' = \frac{1}{\pi}\,\left(\,\frac{a_0}{a}\,\right)^2\, M\,$
reproduces the flat rotation curves.  But
we expect that a more general profile should be of
the form
$
M' = \left[\,\xi\,\left(\,\frac{a_0}{a}\,\right)+\frac{1}{\pi}\,
\left(\,\frac{a_0}{a}\,\right)^2\,\right] \, M\,,
$
with $\xi >0$ which ensures that $M' > 0$ when $a \gg a_0$.\footnote{We have 
neglected terms like $(\,a_0/a\,)^3,\, (\,a_0/a\,)^4, ....,$
because they clearly do not lead to flat rotation curves in the regime 
$a \ll a_0$ and are thus excluded. Therefore, this MDM profile
represents the most general profile. Note added: This discussion has since been
superseded by more recent work by the authors.  See arXiv:1601.00662 
[astro-ph.CO].}
For the more general profile, 
the entropic force expression
is replaced by
$
F = m a_N \left[ 1 + \xi\,\left(\,\frac{a_0}{a}\,\right)+
\frac{1}{\pi} \left( \frac{a_0}{a} \right)^2 \right].
$

Sanders \cite{Sanders1999} studied the virial discrepancy (i.e.,
the discrepancy between the observed mass and the dynamical mass) in 
the contexts of Newtonian dynamics and MOND.  We \cite{Edm1} 
have adapted his approach to the case of MDM. For his work,
Sanders considered 93 X-ray-emitting clusters from the compilation
by White, Jones, and Forman (WJF) \cite{FJW1997}.
He found 
the well-known discrepancy
between the Newtonian dynamical mass ($M_{\textrm{N} }$) and the observed mass 
($M_{\textrm{obs} }$):
$
\left \langle \, \frac{M_\textrm{N}}{ M_{\textrm{obs} }} \,\right \rangle 
\approx 4.4\,.
$
Viewing MOND as a modification of inertia Sanders identified the MOND
dynamical mass $M_{\textrm{MOND}}$ as
$\frac{G \,M_{\textrm{MOND}}} { r_{\textrm{out}}^2} = a \, \mu 
\left(\,\frac{a}{a_c}\,\right)\,$.
With $\mu(x) = x/( 1 + x^2)^{1/2}$ as
the interpolating function, it can be easily shown that
$
M_{\textrm{MOND}} = M_\textrm{N} / \sqrt{1 + 
\left(\,\frac{a_c}{a}\,\right)^2}\,.
$
For the sample clusters, Sanders found 
$\langle M_{\textrm{MOND}} / M_{\textrm{obs}} \rangle \approx 2.1.$

For MDM, the observed (effective) acceleration is given by
$a_{obs} = \sqrt{a^2 + a_0^2} - a_0$.  Using
the more general expression for the MDM profile,
we have $a_{obs} = \frac{G M_{MDM}}{r^2} 
\{ 1 + \xi\,\left(\,
\frac{a_0}{a}\,\right)+ \frac{1}{\pi}\,\left(\,\frac{a_0}{a}\,\right)^2 \}$.
Recalling that $a_{obs} = G M_N / r^2$ for Newtonian dyanmics, we get
$
M_{\textrm{MDM}} = \frac{M_\textrm{N}}{1 + \xi\,
\left(\,\frac{a_0}{a}\,\right)+ 
\frac{1}{\pi}\,\left(\,\frac{a_0}{a}\,\right)^2}\,,
$
for the dynamical mass for MDM.

\begin{figure}[h]%
\begin{center}
\parbox{2.3in}{\includegraphics[angle=0,width=1.8in]{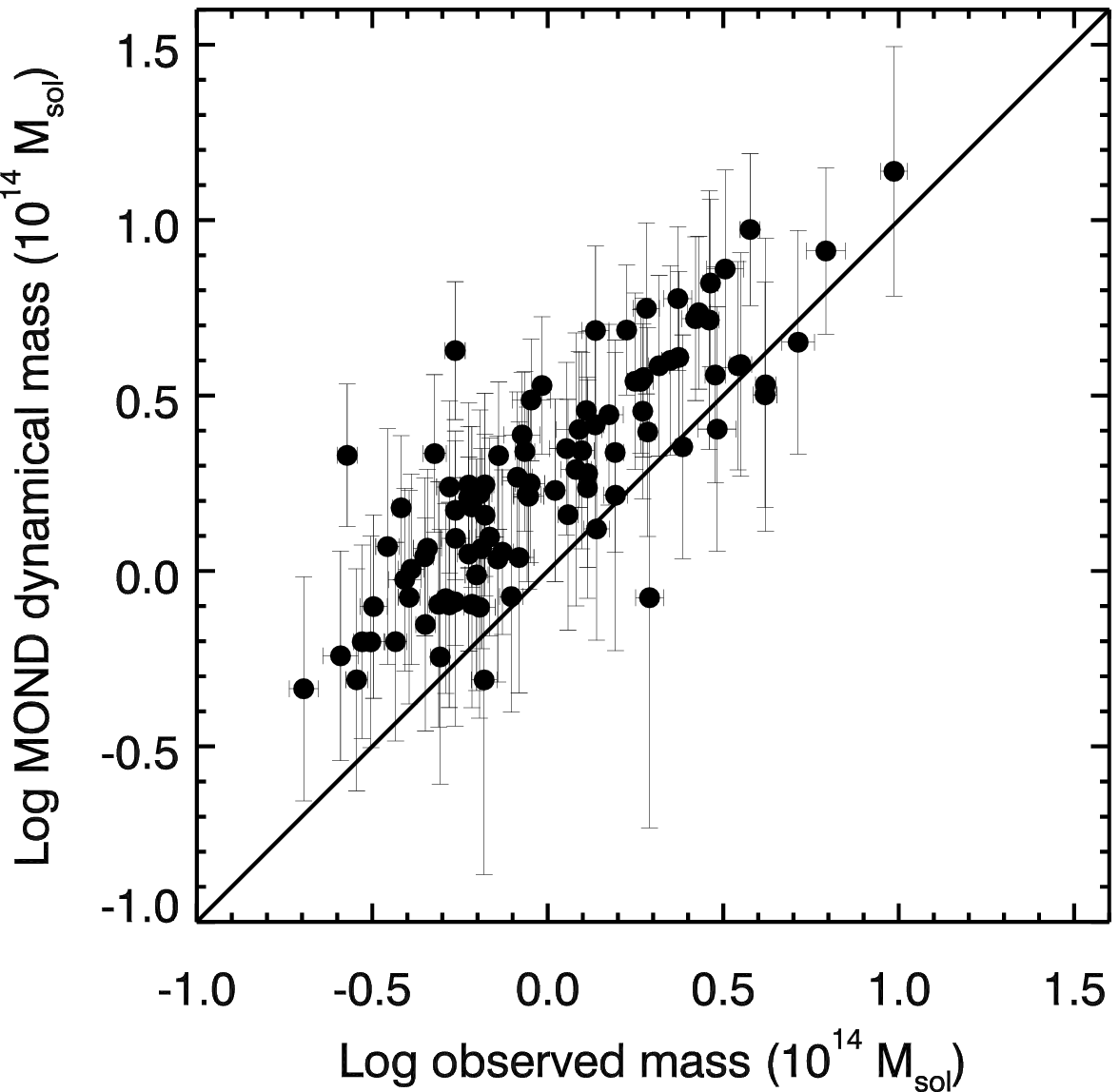}
\figsubcap{a}}
\hspace*{4pt}
\parbox{2.3in}{\includegraphics[angle=0,width=1.8in]{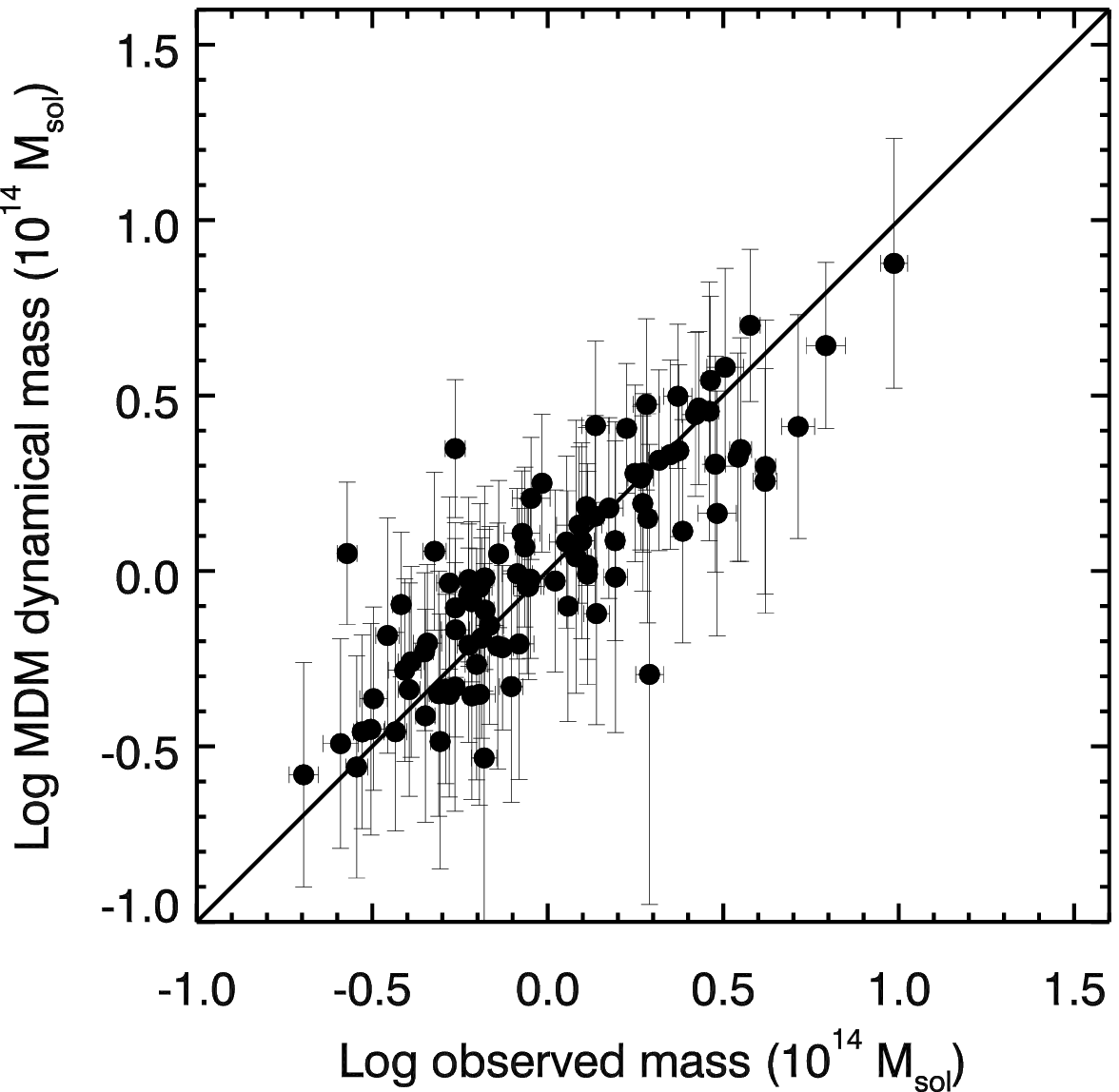}
\figsubcap{b}}
\caption{Fit to galactic cluster data using
(a) MONDian dynamics; (b) MDM dynamics.}
\label{fig5.6}
\end{center}
\end{figure}

In Fig. ~\ref{fig5.6}{a} and Fig. ~\ref{fig5.6}{b}, we show the MOND and MDM 
dynamical masses respectively
against the total observed mass for the 93 sample clusters compiled by
WJF. For MDM
$\xi$ is used as a universal fitting parameter which we find to be 
$\xi \approx 0.5$.
(For completeness we mention that we
have used $\xi = 0$ when fitting galactic rotation curves in the previous 
section. But since now
the galaxy cluster sample in our current study implies $\xi 
\approx 0.5$, we refit the galaxy rotation curves using $\xi 
= 0.5$ and found the fits are nearly identical with a reduction in 
mass-to-light 
ratios of about 35\%.)  To recapture, while Sanders found
$\langle M_{\textrm{MOND}} / M_{\textrm{obs}} \rangle \approx 2.1$, we get
$
\left \langle \, \frac{M_\textrm{MDM}}{ M_{\textrm{obs} }} \,\right \rangle 
\approx 1.0\,.
$
Thus the virial discrepancy is eliminated in the context of MDM! 
At the 
cluster scale, MDM is superior to MOND, as expected.

Finally we comment on strong gravitational lensing in the context of 
MDM and MOND. (Recall that 
strong lensing refers to the formation of multiple images of background 
sources by the central regions of some clusters.)  It is known that
the critical surface density required for strong lensing is
$\Sigma_c = \, \frac{1}{4 \pi} \, \frac{c H_0}{G} \, F(z_l, z_s)$,
with $F \approx 10$ for typical clusters and background sources at 
cosmological distances. 
Sanders \cite{Sanders1999} argued that, in the 
deep MOND limit, $\Sigma_{MOND} \approx a_c / G$.  Recalling that
numerically $a_c \approx c H_0 / 6$, Sanders concluded that
MOND cannot produce strong lensing on its own: $\Sigma_c \approx 5 
\Sigma_{MOND}$.  On the other hand, 
MDM mass distribution appears to be {\it sufficient} for strong lensing
since the natural scale for the critical acceleration for MDM is $a_0 = c H_{0} 
= 2 \pi a_c \approx 6 a_c$, five to six times that for MOND.\\




\end{document}